\documentclass{elsart}
\usepackage{epsfig} 
\usepackage{graphicx}
\begin{document}
\begin{frontmatter}
\title{Performance of prototype BTeV silicon pixel detectors 
       in a high energy pion beam}
\author[Fermi]{J.A. Appel},
\author[Syracuse]{M. Artuso\thanksref{label2}},
\author[Fermi]{J.N. Butler},
\author[Fermi]{G. Cancelo}, 
\author[Fermi]{G. Cardoso}, 
\author[Fermi]{H. Cheung},
\author[Fermi]{G. Chiodini},
\author[Fermi]{D.C. Christian},
\author[INFN]{A. Colautti}, 
\author[Milan]{R. Coluccia},
\author[Milan]{M. Di Corato},
\author[Fermi]{E.E. Gottschalk}, 
\author[Fermi]{B.K. Hall}, 
\author[Fermi]{J. Hoff}, 
\author[Fermi]{P. A. Kasper},
\author[Fermi]{R. Kutschke}, 
\author[Fermi]{S.W. Kwan}, 
\author[Fermi]{A. Mekkaoui},
\author[INFN]{D. Menasce}, 
\author[Iowa]{C. Newsom}, 
\author[INFN]{S. Sala},
\author[Fermi]{R. Yarema},
\author[Syracuse]{J.C. Wang}, and
\author[Fermi]{S. Zimmermann}
\thanks[label2]{Corresponding Author: e-mail:artuso@physics.syr.edu}
\address[Fermi]{Fermi National Accelerator Laboratory, Batavia, IL 60510, USA}
\address[Iowa]{University of Iowa, Iowa City, IA 52242, USA}
\address[INFN]{Sezione INFN di Milano, via Celoria 16 - 20133 Milano, Italy}
\address[Milan]{Universit\'{a} di Milano, Dipartimento di Fisica, 
via Celoria 16 - 20133 Milano, Italy}
\address[Syracuse]{Syracuse University, Syracuse, NY 1344-1130, USA}

\begin{abstract}
The silicon pixel vertex detector is a key element of the BTeV spectrometer. 
Sensors bump-bonded to prototype front-end devices 
were tested in a high energy pion beam at 
Fermilab. 
The spatial resolution and occupancies as a 
function of the pion incident angle were measured for various sensor-readout combinations. The data are 
compared with predictions from our Monte Carlo simulation and very good
agreement is 
found.

\end{abstract}
\end{frontmatter}

\section{Introduction}
BTeV is an experiment expected to run in the new Tevatron C0 interaction
region at Fermilab in $\approx$ 2006. It is designed to perform precision
studies of $b$ and $c$ quark decays, with particular emphasis on mixing, CP
violation, rare and 
forbidden decays ~\cite{BTeVproposal}. This experiment takes advantage of two
important features 
of the ``forward'' region: the correlation in the direction of the
produced $b$ and $\bar{b}$, that improves the flavor tagging efficiency, and the
boost that is exploited in our trigger algorithm based
upon the identification of detached charm and beauty decay vertices. The unique
feature of BTeV is that this algorithm is implemented in the first trigger
level~\cite{BTeVtrigger}. 
Consequently, the vertex detector must
have a fast readout, superior pattern recognition power, 
small track extrapolation errors,
and good performance even after high radiation dose.
Silicon pixel sensors were chosen because they
provide very accurate space point information and
have intrinsically low noise and high radiation hardness.

In this paper we report the results of the 1999-2000 BTeV silicon pixel detector 
beam test and we compare them to Monte Carlo predictions.
The beam test was carried out at Fermilab in a 227 GeV/c pion test beam.
The pixel detectors tested were hybrid assemblies
of several combinations of pixel readout chip prototypes
developed at Fermilab, and single-chip sensor prototypes. 
The main goal of our studies was to measure
the spatial resolution attainable along
the short pixel dimension for different
sensor technologies and for different readout electronics configurations.
In particular, we have performed an extensive investigation
of the effects of varying the discriminator 
threshold and the front-end device digitization precision.

\section{Experimental setup}
\begin{figure}
\begin{center}
\epsfig{figure=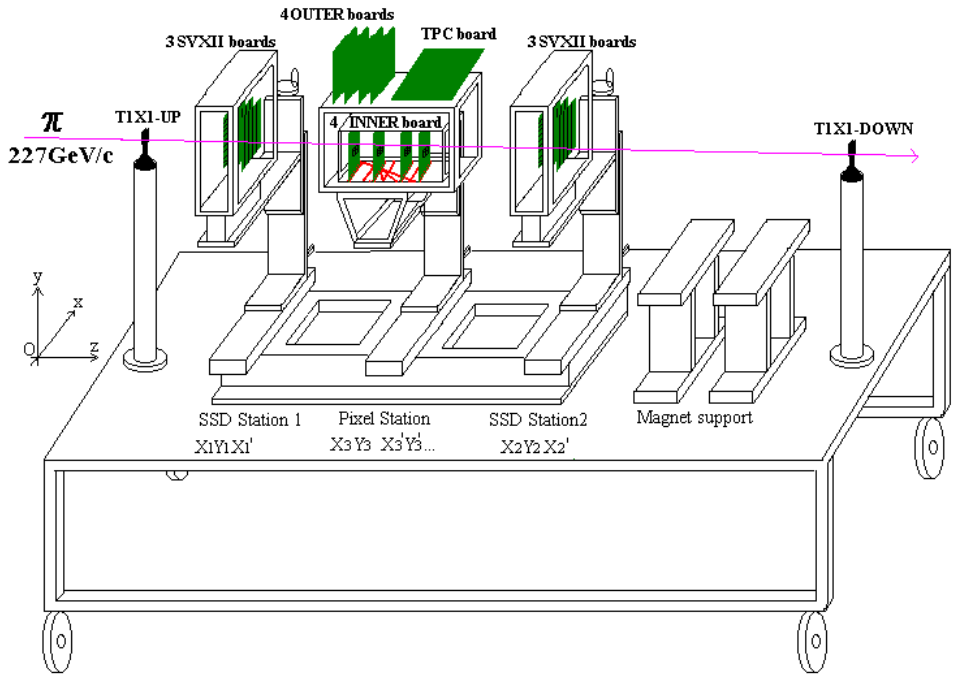,width=5in}
\end{center}
\caption{Schematic drawing of the silicon telescope and 
data acquisition system installed in the thermo-controlled hut.
\label{telescope}}
\end{figure}

The data were collected at the
MTest beam line located in the Meson Area at Fermilab.
Fig.~\ref{telescope} shows the experimental set-up. The pixel devices 
were located between two stations of silicon microstrip detectors (SSD's)
that provided tracking information to an accuracy of about 2 $\mu$m in the
$x$ direction, corresponding to the ``small pixel dimension'' (50 $\mu$m). The
pixel hybrid devices were mounted on printed circuits boards, held inside
an aluminum box, where their location was determined by precision machined
slots. One of the pixel devices in the telescope could be positioned
in slots at various angles with respect to the beam direction.
This enabled us to measure the properties of various pixel prototypes
as a function of the pion incident angle.
\noindent

\begin{figure}
\begin{center}
\epsfig{figure=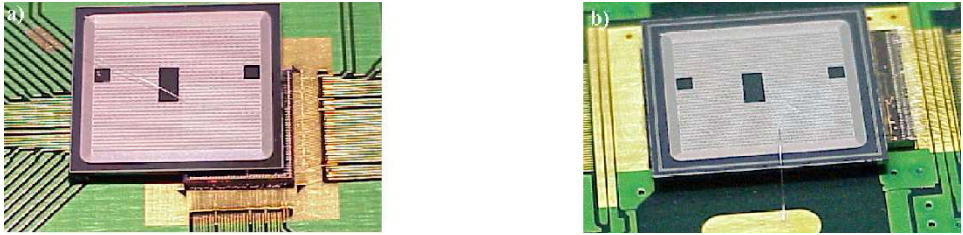,width=5in}
\end{center}
\caption{The photograph on the left shows CiS ``ST1'' bonded to FPIX0.  
The sensor is on top in the photograph.  Only pixels
in the lower right corner of the sensor are bonded to the FPIX0, which is 
smaller than
the sensor.
The photograph on the right shows SII ``ST1'' bonded to FPIX1.  
All sensor channels are bonded to readout pixels.
\label{FPIX0AND1}}
\end{figure}

The pixel detectors tested are all from the ``ATLAS prototype submission'' 
\cite{ATLAS_1st}, and all have 50 $\mu$m $\times$ 400 $\mu$m pixels. They use
the $n^+/n/p^+$ technology, with the pixel electrodes located on the ohmic
side of the device. In order to achieve good inter-pixel insulation, two
approaches have been tried. The ``$p$-stop'' technique uses $p^+$ implants
between the pixel cells, deposited through a mask with the chosen implant
geometry, whereas  in the ``$p$-spray'' technique, pixel implants are deposited
after a shallow $p^+$ layer is deposited uniformly throughout the active area.
Two of the sensors were produced by CiS, Germany, 
and the other three by SII, Japan.
Fig.~\ref{FPIX0AND1} on the left shows an FPIX0 bonded to the CiS $p$-stop 
sensor. The instrumented portion of the sensor is 11 columns $\times$ 64
rows. The CiS sensors (one $p$-stop ST1 and one $p$-spray
 ST2) were indium bump bonded to FPIX0 readout chips by 
Boeing North America, Inc. 
The SII sensors (two $p$-stop ST1's and one $p$-spray ST2) were indium bump 
bonded to FPIX1 readout chips by Advanced Interconnect Technology Ltd. (AIT,
Hong Kong).
The depletion voltage, measured through the dependence of the leakage current
on the reverse bias applied, is 85 V for the CiS sensors and 45V for the
SII sensors. The sensor thickness was about 300$\mu$m.

The front-end device to be coupled to the pixel sensors must satisfy several
challenging requirements. In order to obtain the optimal spatial
resolution achievable with the chosen pitch, analog information is needed. On
the other hand, it is necessary to transfer the hit information very quickly
to the trigger processor in order to be able to use it in the first trigger
level algorithm. The development effort towards the final chip
~\cite{Fpix} has proceeded along several steps of increasing complexity.

The first iteration, FPIX0, has 12 columns of 64 rows.
Each FPIX0 readout pixel contains an amplifier, a comparator, and a peak sensing
circuit.  When any comparator fires, a FAST-OR signal is asserted.  FPIX0
provides a zero-suppressed readout of hit pixels.  The information read out
consists of hit row and column numbers, together with a voltage level which is
proportional to the peak pulse height. For these measurements, the analog 
output was digitized by an external 8-bit flash ADC.

FPIX1 is the second-generation pixel readout chip developed at Fermilab, 
and is the first implementation of a high speed readout architecture.
Each FPIX1 cell
includes 4 comparators: one comparator is used to provide sparse readout and
the other three implement a 2 bit flash ADC for each cell.

Two scintillation counters located upstream and downstream of the telescope
determined the trigger, through a coincidence with the FAST-OR signal from
the FPIX0 $p$-stop hybrid detector. 
The apparatus was located inside a temperature controlled
enclosure, maintained 
at $\approx 20^\circ$C.\hfill

\section{Pixel charge calibration}
A relative calibration of the ADC response of each cell on a readout chip
has been performed by injecting charge into
individual pixels, sending a voltage pulse to 
a calibration capacitor in each front-end channel. With 
this method the gain and equivalent noise
charge are determined up to a scale factor associated with the value of
the input capacitor. In order to perform an
 absolute calibration, we have used two x-ray sources
(Tb and Ag foils excited by an Am $\alpha$ emitter). The X-rays produce
known signals in the sensors and lead to an absolute
determination of the electronic gain as well as the equivalent noise charge
and discriminator thresholds.

\begin{figure}
\begin{center}
\begin{tabular}[c]{cc}
    \epsfig{figure=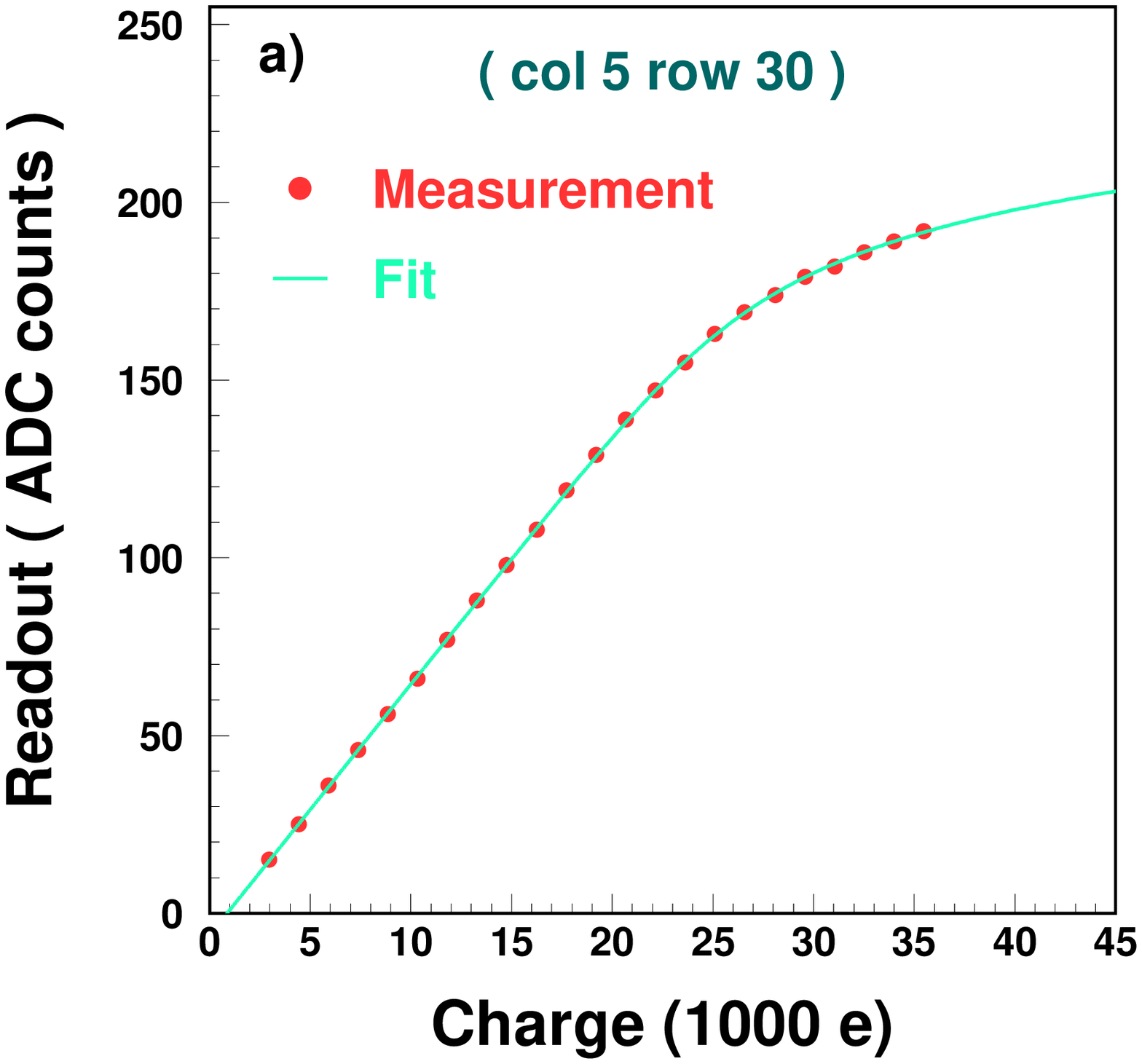, height=2.8in} &
    \epsfig{figure=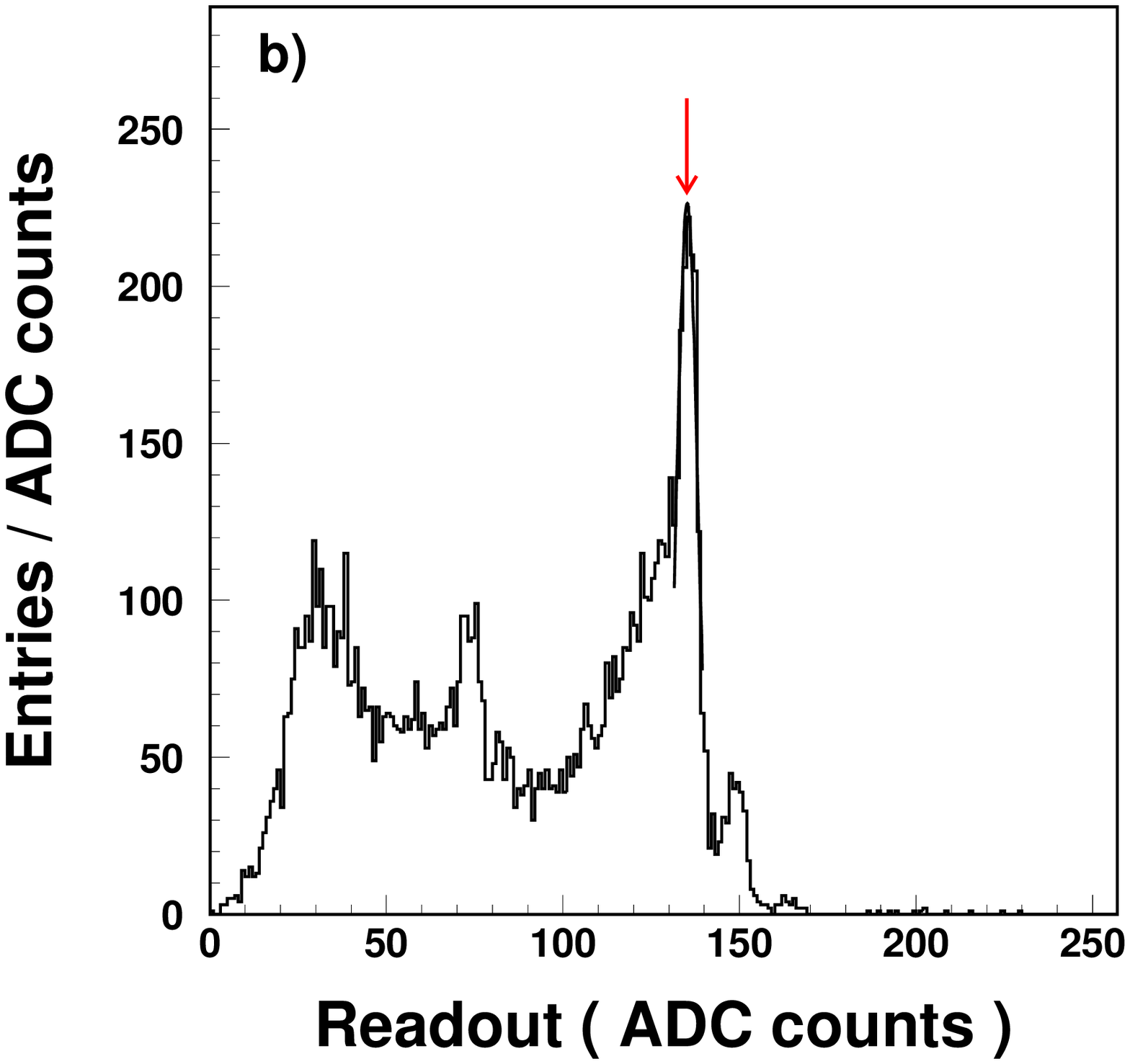, height=2.8in}
\end{tabular}
\end{center}
\caption{Calibration for the FPIX0 $p$-spray (CiS) sensor.  
(a) single channel calibration curve with
a nonlinear fit superimposed;
(b) Tb x-ray spectrum.
The arrow indicates the K$_\alpha$ peak. 
\label{calfpix0} }
\end{figure}

In Fig.~\ref{calfpix0}, the pedestal subtracted
and gain equalized spectrum of a Tb X-ray source 
measured for an FPIX0-instrumented detector is shown.
The single channel calibration curve fits
the data points very well. FPIX0 contains two different input cells
characterized by different gains (
``high gain'' and ``standard gain'').
In the following front-end
characterization, the quoted central value gives the average quantity and the
uncertainty gives the distribution rms within a chip.
For most of the data taking, the discriminator threshold for the FPIX0 $p$-stop
was set to a voltage
equivalent to  2500$\pm$400 e$^{-}$ for the standard gain cells, and
1500$\pm$230 e$^{-}$ for the high gain cells. 
For the FPIX0 $p$-spray device the
corresponding thresholds were typically 2200$\pm$350 e$^{-}$ and 1250$\pm$160
e$^{-}$. The equivalent noise charge is 105$\pm$15 e$^{-}$ for standard
gain cells, and 83$\pm$15 e$^{-}$ for for high gain cells of the FPIX0
$p$-stop hybrid pixel devices.  The corresponding values for the FPIX0
$p$-spray devices are 80$\pm$10 e$^{-}$ for standard gain cells, 
and 67$\pm$8  e$^{-}$ for  high gain
cells.  An additional contribution due to the external
buffer amplifier and ADC of about 400$\pm$150 e$^{-}$ for FPIX0 standard gain
cells and 205$\pm$95 e$^{-}$ for FPIX0 high gain cells was present.

\begin{figure}
\begin{center}
\epsfig{figure=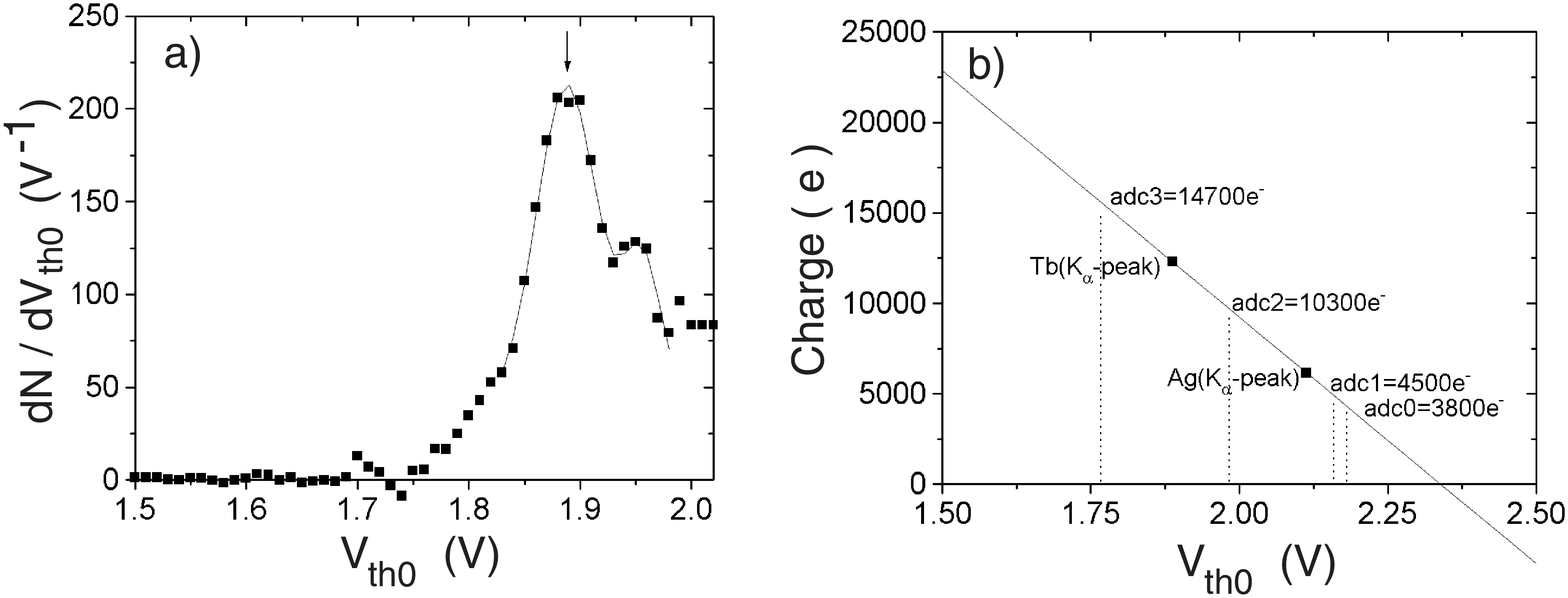,width=5in}
\end{center}
\caption{Calibration for the FPIX1 $p$-stop (SII) sensor:  
(a) Tb x-ray differential spectrum versus the
external voltage $V_{th0}$.  The arrow indicates the K$_\alpha$ peak; 
(b) threshold calibration curve 
obtained from the Tb and Ag K$_\alpha$ peaks and
the four FPIX1 thresholds in nominal operating conditions.
\label{calfpix1} }
\end{figure}

The absolute calibration of the FPIX1 readout chip response 
was determined by measuring the differential 
counting rate due to the same two X-ray sources.
In Fig.~\ref{calfpix1} the differential 
counting rate obtained by sweeping the external voltage
V$_{th0}$ for a hybrid detector exposed to a Tb source is shown. 
The peak of the derivative of the efficiency curve 
defines the  threshold voltage V$_{th}$ 
corresponding to the known K$_{\alpha}$ line.
The procedure is repeated with a Ag source with a line producing a pulse
height within the linearity range of FPIX1 to make an absolute measurement of
the slope of the flash ADC response.
The threshold used 
for FPIX1 $p$-stop hybrid detector is
3,800 $e^{-}$. In addition, the three thresholds determining the digitization
bin sizes of the flash ADC were 4,500 e$^{-}$, 10,300 e$^{-}$, and 14,700 e$^{-}$.
The threshold dispersion for each discriminator 
was fitted to a Gaussian distribution.
The standard deviation was $\approx 380\ e^{-}$ in each case.
The front-end equivalent noise charge is
110$\pm$30 e$^{-}$, where the quoted uncertainty reflects the rms 
spread within a chip.

\section{Monte Carlo simulation of the pixel sensor performance}
In order to identify the sensor and readout properties that 
are optimal for our
experiment,
we have developed a Monte Carlo simulation~\cite{marina}
including detail modeling of 
all the main physical processes affecting the development and
collection of the signal in silicon.  

Electrons and holes 
are produced by the energy deposited in the sensor 
by the charged track, and are simulated  
including excitations, low energy ionization~\cite{geant3}, and energetic
knock-on electron ($\delta$ ray) emission~\cite{delta-ray}.
The algorithm will be discussed referring to ``$n^+/n/p^+$'' pixel
sensors,
where
the relevant charge carriers are electrons.
The drift motion of the electrons along the $\vec{E}$ direction is described 
by the current density equation:
\begin{equation}
\vec{J}_e = - q\rho _e\mu _e\vec{E}\\
\end{equation}
where $q$ is the magnitude of the electron charge, $\mu _e$ 
is the electron mobility and $\rho _e$ is the number of free electrons
 per unit volume.
The speed is related to the electric field 
through the drift mobility $\mu _e$. An experimental $\mu _e$
parameterization~\cite{cms} is adopted, including non-linear effects at high
fields. 

The charge cloud spreads laterally due to diffusion. The 
parameter 
characterizing  the drift in the 
electric field ($\mu _e$) is related to the parameter describing 
the diffusion of the charge cloud 
($D_e$) by the Einstein equation:
\begin{equation}
D_{e}=\frac{kT}{q} \mu_{e}
\end{equation}
where $D_e$ is the electron diffusion coefficient and $kT$ is 
the product of the
Boltzmann constant and the absolute temperature of the silicon.
The average square deviation with respect to the trajectory of the collected 
charge without diffusion is $<\Delta r^2>=2<D>\Delta t$. Note that in
$n^+/n/p^+$ sensors, the region of highest 
electric field is the farthest from
the
signal electrodes. This is the region characterized by lower mobility
and, consequently, lower diffusion coefficient. However the collection time
is also influenced by the non-linear behavior of the drift velocity, thus the
diffusion radius is less sensitive to this effect.

\begin{figure}
\begin{center}
\epsfig{figure=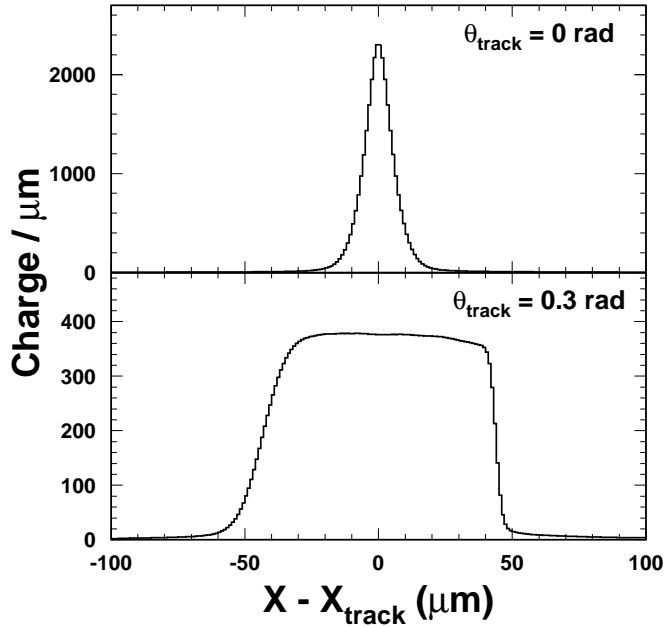,width=4in}
\vspace{-0.5cm}
\end{center}
\caption{Charge distribution for a track angle of  0 rad (top) and  0.3 rad 
(bottom) with respect to the normal to the detector plane.
\label{chdist} }
\end{figure}

Fig.~\ref{chdist} shows the charge spread for track angles of 0 rad
and 0.3 rad with respect to the normal to the detector plane. Diffusion 
determines the shape of the charge distribution at
small incident angles and allows interpolation between pixel centers using charge
weighting. At larger angles the charge division is linear and is dominated
by the cluster broadening 
produced by the track inclination. The spatial resolution is
determined by the precision of the charge measurement and by the pixel pitch. 
Our pixel detectors are expected to be very low-noise devices, as can be
inferred by the data discussed previously.
In order to take full advantage of
this feature, the 
discriminator threshold spread 
needs to be small, as this spread needs to be added in quadrature to the
intrinsic noise to determine the overall noise performance of the system. 
As mentioned before, the threshold spread measured in the
devices used in the test beam was about 380 $e^-$. Therefore noise and 
threshold spread figures
are not limiting factors in the detector performance.

\section{Data analysis and results}
\subsection{Analysis Method}
\label{monitor}
The track position at each plane in the telescope is reconstructed using analog
charge weighting. 
Each track is fitted to a straight line using the Kalman filter
technique and tracks with a good $ \chi ^{2}$  are retained for further
analysis. The Kalman fitter includes information from strip detectors and 
additional pixel sensors, but excludes the device under test. The track 
parameters  allow the determination of the
position ($x_T,y_T$) at which the beam intersects the sensor being
characterized. Although the pixel detectors measure two coordinates, the
discussion will focus on the resolution that can be achieved in the direction
with smaller pitch, $x$.

In order to  predict the track impact
point $x_T$, it is important to align the
individual detectors in the telescope. We use a right-handed coordinate system
$x,y,z$, where $z$ corresponds to the beam direction, 
$y$ is oriented along the vertical direction and $x$ is the horizontal
axis. Each plane is 
defined by 3 offset parameters $\delta
x$, $\delta y$ and $\delta z$ and three angles $\alpha$, $\beta$ and $\phi$
defining their orientation through three rotations around the $x$, $y$ and
$z$ axes respectively. We have developed two alignment procedures, implemented using the
first 1000 events in a run. A ``manual'' technique
finds the optimal values of some parameters, while maintaining 
geometrical parameters that are well measured at their known value. 
In a first iteration, 
translational offsets are determined such that the residual
distributions are centered around zero. Next, the two relevant
pixel plane angles, $\beta$ and $\phi$ are determined by minimizing the
residual widths in the $x$ and $y$ directions. A second, automatic 
iterative alignment procedure has been 
developed performing a global minimization of the $\chi
^2$ of the fitted tracks. This procedure 
guarantees that all the runs are aligned in a consistent fashion.

The $x$ coordinate measured in the pixel plane is determined in two steps. 
First we obtain the
``digital'' coordinate:  
\begin{equation}
x_{D}=\frac{x_{R}+x_{L}}{2},
\label{aldig}
\end{equation}
where x$_L$ (x$_R$) is the local coordinate of the left-most (right-most) hit
in the cluster. The ``left'' pixel is the one with the smallest $x$ coordinate.

Subsequently we refine this measurement with an empirical correction 
term expressed as a function of 
the variable $\eta$ \cite{Turchetta} defined as:
\begin{equation}
\eta=\frac{q_{R}-q_{L}}{q_{R}+q_{L}},                   
\label{eta}
\end{equation}
where q$_{R}$ and q$_{L}$ are the charges deposited in the pixels located at
the right and left boundaries of the clusters. 
Thus the measured position is given by $x_p=x_{D}+f(\eta)$.                    
The correction $f(\eta)$ 
is a function of the number of
pixels in the cluster  and the track angle, and is determined by fitting an
independent data set.

The measured spatial resolution of a pixel plane is the difference
between the reconstructed pixel position $x_p$ and the track position $x_{T}$,
known with an accuracy of about 2
$\mu$m, limited by multiple Coulomb scattering from the material in the
telescope and the intrinsic hit resolution of the various tracking stations.
This prediction uncertainty is subtracted in quadrature from the measured spatial 
resolution in the results shown below.

\subsection{Charge signal distributions}
\begin{figure}
\begin{center}
\epsfig{figure=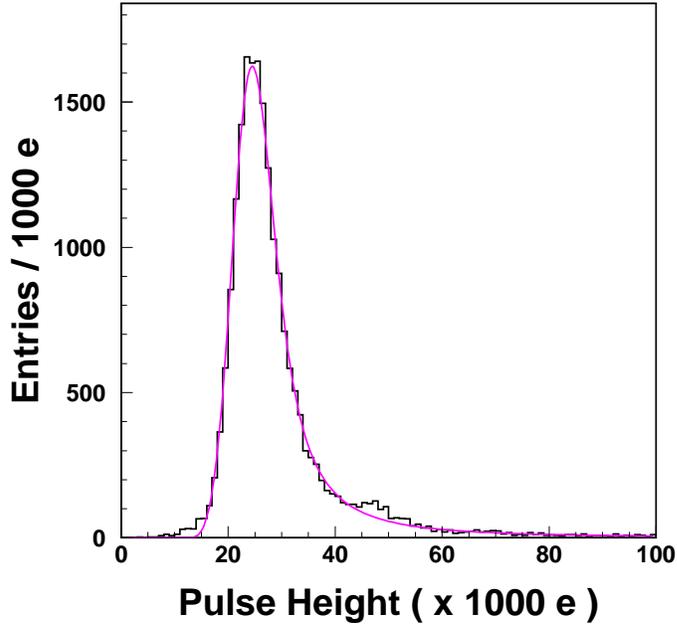,width=4in}
\end{center}
\caption{0$^{\circ}$ track pulse height distribution for a
CiS $p$-stop sensor bump-bonded to an FPIX0 readout chip. 
The Blunck-Leisegang curve \cite{bl} fit is superimposed on the plot. 
\label{Landauonecell} }
\end{figure}

The 8 bit analog information provided by the FPIX0-based readout electronics
 allowed us to measure
several features of the charge signal in various sensors.
The measured pulse height distributions are fitted using
a Landau function convoluted with a Gaussian ~\cite{bl}:
\begin{equation}
f(E)=N\int_{-\infty}^{+\infty}dE'
\frac{e^{-\frac{(E-E')^{2}}{2\sigma_{g}^{2}}}}{\sqrt{2\pi\sigma_{g}^2}}
\frac{\phi(\frac{E'-E_{mp}}{\xi}+\lambda_{0})}{\xi}
\end{equation}
\noindent
where N is a normalization constant, $\frac{\phi(E')}{\xi}$ is the 
Landau probability distribution, with $\lambda_{0}=-0.223$ and
$E_{mp}$ is the most probable energy loss.
For 300 $\mu$$m$ of silicon we have $\xi \approx$ 5.35 KeV,
E$_{mp} \approx$ 84.02 KeV, and $\sigma_{g} \approx$ 5.95 KeV.
The study of the charge collected in single
pixel clusters
for tracks at normal incidence determines the straggling function in our
sensor. This can be compared with model expectations and data from silicon
strip sensors.
The fit includes $N/\xi$, $E_{mp}+\lambda_{0}$, $\xi$, $\sigma_{g}$ as 
free parameters.
Fig.~\ref{Landauonecell} shows the pulse height distribution
for an FPIX0 $p$-stop sensor. The full width at half maximum (FWHM) is 
10,000 $e^-$.

\begin{figure}
\begin{center}
\epsfig{figure=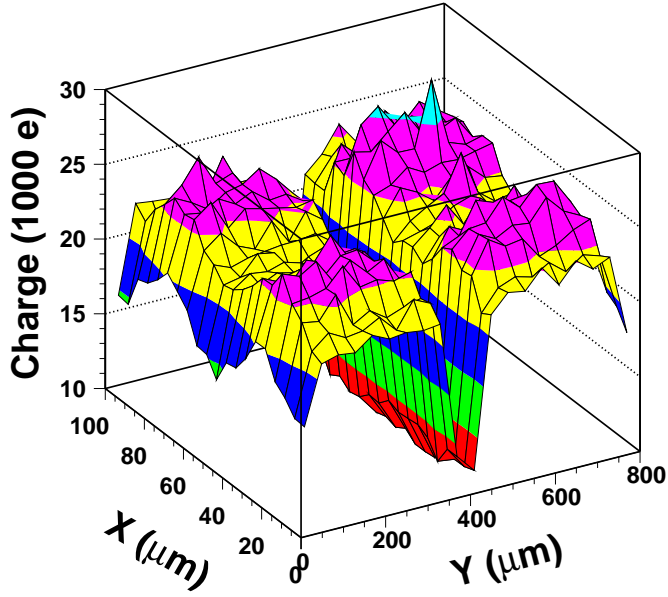,width=4in}
\end{center}
\caption{
Average pulse height versus track position for the  ST2 CiS $p$-spray 
sensor bump-bonded to an FPIX0 readout chip.
\label{2dchargeloss}}
\end{figure}

The agreement with the parameters of the theory and with previous
measurements in strip detectors \cite{Hancock} demonstrates that we are achieving full
charge collection for $p$-stop sensors. We can use the mapping of the average 
pulse height as a
function of the track position to single out regions with poorer charge
collection properties.
Fig. ~\ref{2dchargeloss}
shows the average pulse height as a function of the track position 
for an FPIX0 $p$-spray hybrid detector.
There is an obvious dip at the boundary between two pixels in the $y$
direction, indicating a strong charge 
collection inefficiency. A less pronounced dip at the interpixel boundary in
the $x$ direction is also present.
This collection deficit is
conjectured to be induced by a parasitic path to the
bias grid, introduced to apply reverse bias to all the pixel
cells during wafer testing of the devices  ~\cite{Alam}.
An improved design of $p$-spray sensors has been developed, and is expected to
overcome this limitation \cite{rohe2}.

Table~\ref{tablelandau} summarizes the FWHM and the parameters 
of the Blunck-Leisegang function
for various incident track angles.  
\begin{table}
\centering
\begin{tabular}{|l|l|l|l|l|l|} \hline
{\bf Angle[deg]} & {\bf $<$E$>$/E$_i$} & {\bf E$_{mp}$/E$_i$}
& {\bf $\xi$/E$_i$} & {\bf FWHM/E$_i$} & {\bf $\sigma_g$/E$_i$}\\ \hline
0  & 28900 & 23400$\pm$28  & 1460$\pm$28 & 9600  & 2790$\pm$53  \\ 
\hline
5  & 28400 & 22800$\pm$44  & 1520$\pm$32 & 9610  & 2710$\pm$61  \\ 
\hline
10 & 29600 & 24000$\pm$43  & 1500$\pm$31 & 10400 & 3250$\pm$60  \\ 
\hline
15 & 30200 & 24450$\pm$45  & 1610$\pm$34 & 10670 & 2930$\pm$63  \\ 
\hline
20 & 31800 & 26050$\pm$47  & 1650$\pm$36 & 10680 & 3300$\pm$72  \\ 
\hline
30 & 35000 & 39300$\pm$64  & 1660$\pm$50 & 12800 & 4000$\pm$77  \\ 
\hline
%
\end{tabular}
\caption{ Parameters of the Blunck-Leisegang function and FWHM of the
 collected charge distribution for FPIX0 $p$-stop detector for various
 incident track angles.  \label{tablelandau} }
\end{table}

\subsection{Pixel occupancy}

\begin{figure}
\begin{center}
\begin{tabular}[t]{cc}
   \epsfig{figure=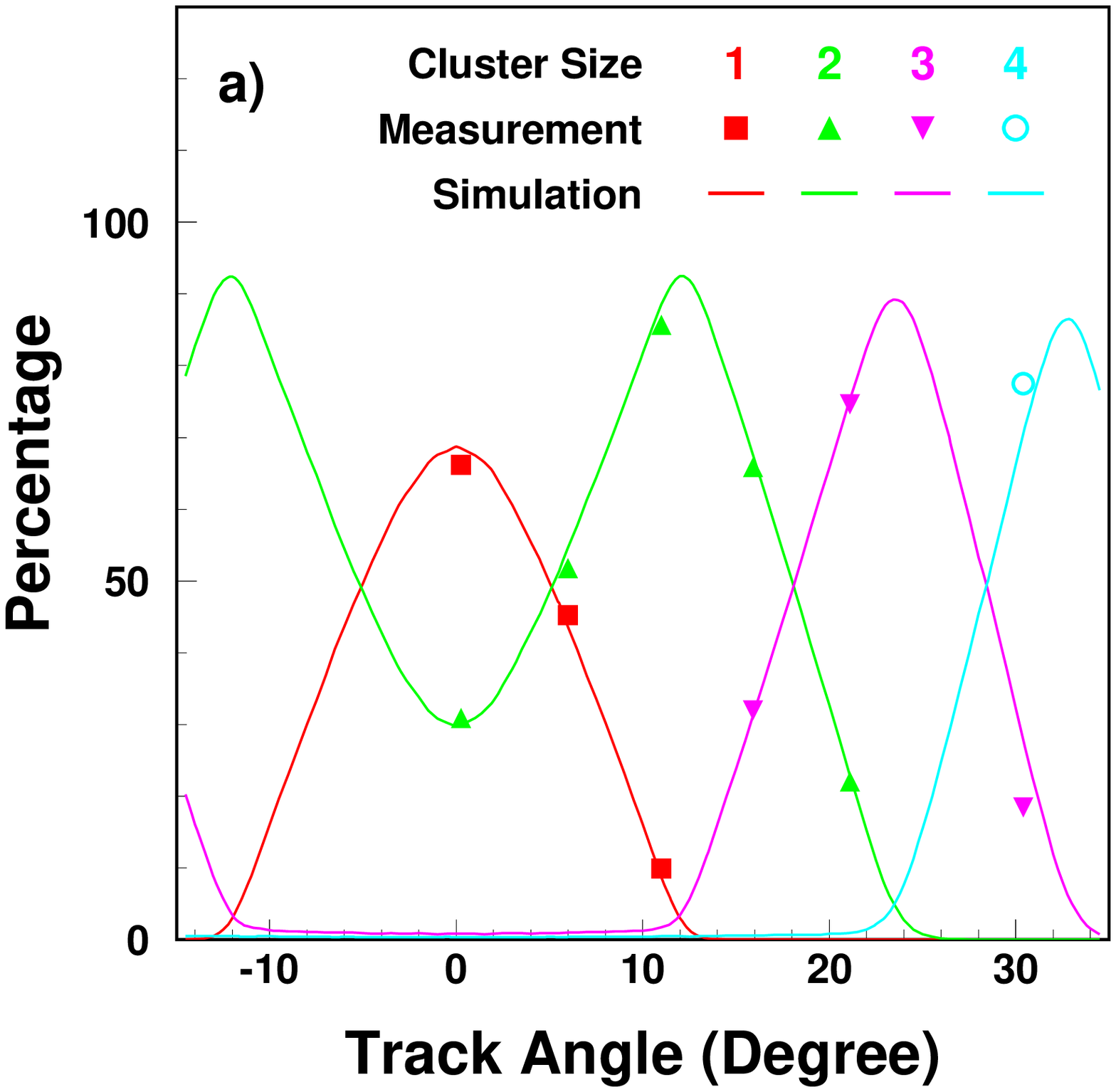,width=2.5in}
   \epsfig{figure=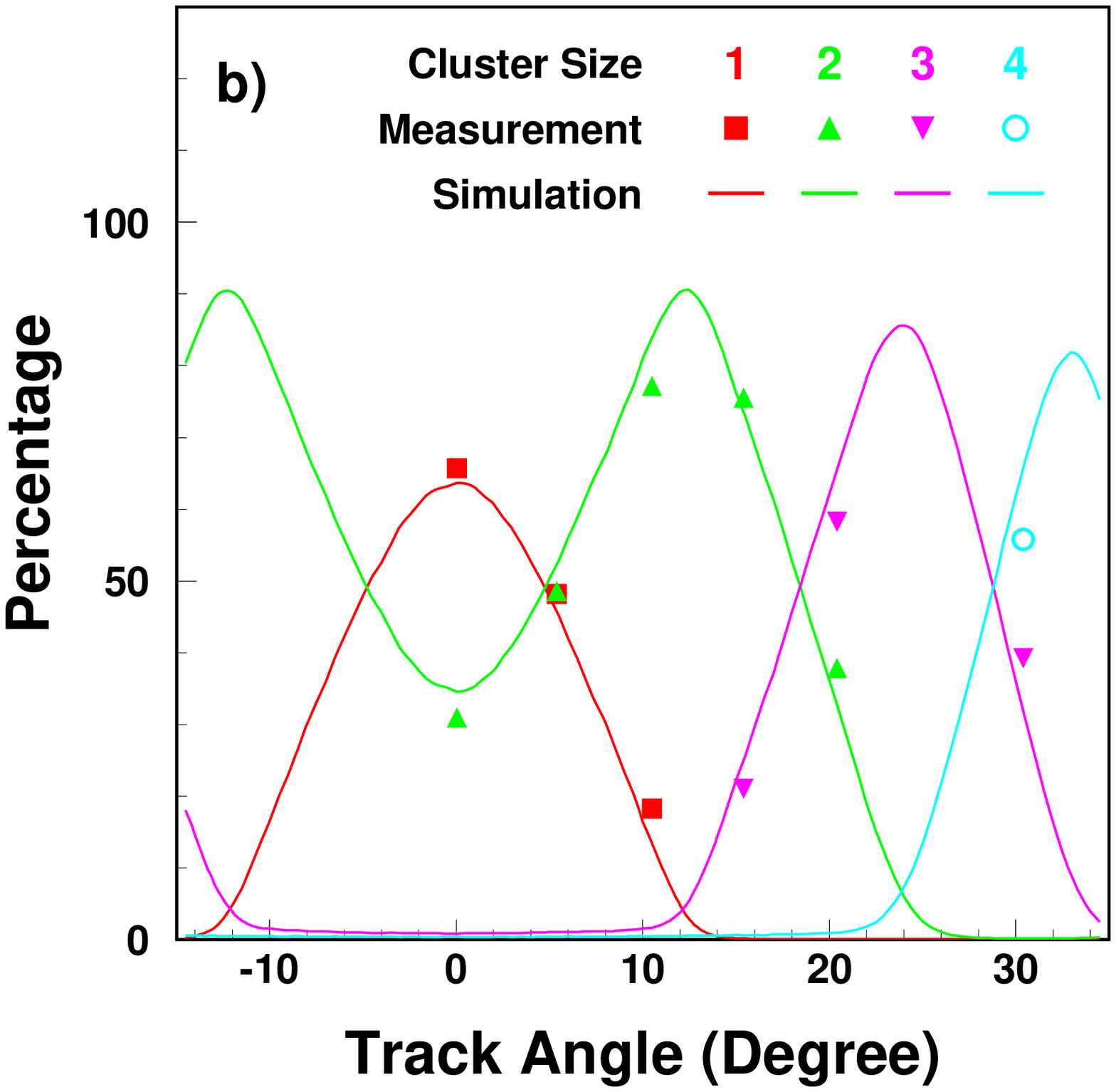,width=2.5in}
\end{tabular}
\end{center}
\caption{ Cluster row multiplicity fractions for various angles 
of incidence. The plot on the left shows data from FPIX0 $p$-stop
($V_{bias}=-140 V$ and $Q_{thr}=2500e^{-}$), the plot on the
right from FPIX1 $p$-stop ($V_{bias}=-75 V$ and $Q_{thr}=3700e^{-}$).
The curves are our Monte Carlo predictions.
\label{Sharingglobal}}
\end{figure}

We performed a thorough study of the ``row occupancy'', namely the
number of $x$ pixels in a cluster as a function of the track incident angle
and the sensor and electronics operating conditions.
Fig.~\ref{Sharingglobal}a shows the number of rows in a cluster as a function
of the angle measured for FPIX0 sensors in nominal 
operating conditions as well as our Monte Carlo predictions.
The agreement between data and simulation is
excellent over the whole angular range studied.  
The FPIX1 based hybrid detectors have been studied with the same procedure,
and the results are shown in
Fig. ~\ref{Sharingglobal}b.

\subsection{Position resolution}
\label{Positionresolution}

We studied the spatial resolution achieved with different
sensor-readout electronics combinations for 
six track incident angles (0, 5, 10, 15, 20 and 30 degrees). 
The  bias voltage  was 140 V for CiS devices and 75 V for SII devices,
corresponding to overdepleted sensors.
The data sample was about
50,000 events per measurement.
We also studied the
sensitivity of our results to key parameters such as 
the bias voltage applied to the sensor and the discriminator
threshold. We collected
about 10000 events for each operation condition.

Only events containing a single reconstructed track, 
characterized by a cluster including at least two strips 
in the most upstream and downstream $x$-measuring SSD planes are
used in this analysis.

\begin{figure}
\begin{center}
\epsfig{figure=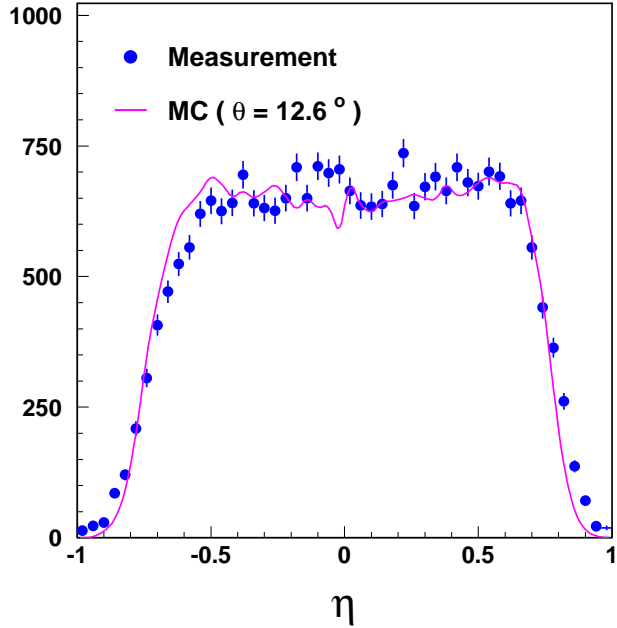,width=4in}
\end{center}
\caption{
The curve shows the predicted $\eta$ distributions for 12.6$^{\circ}$
track angle and 2 pixel clusters. 
The measured points are with a FPIX0-CiS $p$-stop hybrid detector.
\label{etadis} }
\end{figure}

Fig.~\ref{etadis} shows the measured $\eta$ distribution for 10$^{\circ}$ 
tracks. An entry is made in these histograms only if
adjacent pixels in a single column are hit. The curve shows 
Monte Carlo predictions for a track angle offset by 1.8$^{\circ}$
with respect to the value determined with the automatic alignment procedure
described before. The 10$^{\circ}$ angle is chosen 
to determine the global offset reflecting a rotation of the beam
axis with respect to our apparatus. This is
because at this angle most of the tracks form 2 pixel
clusters and thus the 2 pixel $\eta$ distribution has the best statistical
power and can be predicted more accurately.
The agreement between Monte Carlo and data
is excellent, indicating that the various factors influencing the charge
sharing are well modeled. 

\begin{figure}
\begin{center}
\epsfig{figure=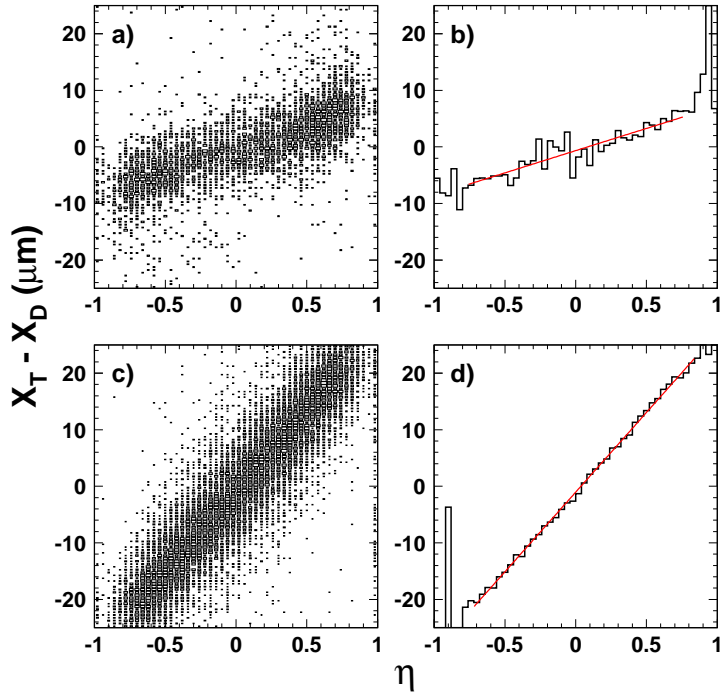,width=4in}
\end{center}
\caption{
Scatter plots of the digital residual versus the $\eta$: (a) for 0$^{\circ}$
and
(c) for 10$^{\circ}$ track angles; (b) and (d) show the corresponding corrections $f(\eta
)$
extracted from a linear fit to the average of the digital residual distributions. 
These data correspond to a FPIX0-CiS $p$-stop hybrid detector.
\label{etafun} }
\end{figure}

The empirical correction $f(\eta)$ is determined by requiring that the average
value of the residuals of the $x_D$ distribution is 0. The average residuals
as a function of $\eta$ can be fitted with a variety of functions, the
simplest one being a straight line. In the FPIX1 case, it is convenient to
determine a correction for each of the 16 possible values of the pair $(q_L,q_R)$.
Fig.~\ref{etafun}.a shows the difference $\Delta x_D$ 
between $x_T$ and the digital centroid $x_D$ as a function of $\eta$ for
tracks at normal incidence;
Fig.~\ref{etafun}.b shows the linear fit to the residual average
distribution $f(\eta)$; Fig.~\ref{etafun}.c and d show the corresponding
plots for at 10$^{\circ}$ incident angle.  
We have determined a set of $f(\eta,n_{pix},\theta_T)$'s
using similar plots for each detector.
This is done using independent track samples,
dividing $\eta$ into 50 sub-intervals, and for each sub-interval determining
the average value of $\Delta x_D$.

\begin{figure}
\begin{center}
\epsfig{figure=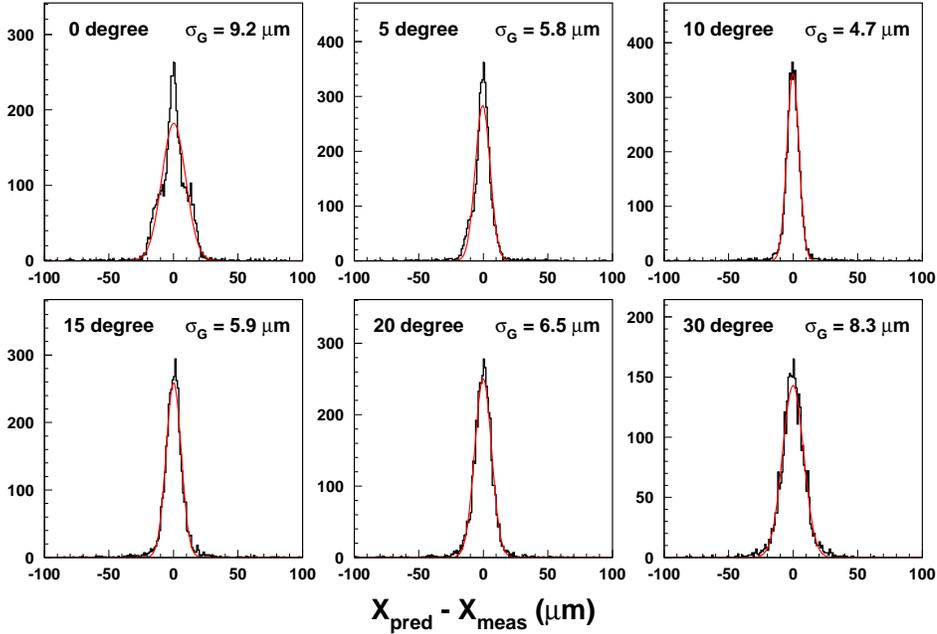,width=5in}
\end{center}
\caption{Residual distributions for the FPIX0 $p$-stop detector. $\sigma_G$ is the 
         standard deviation of the Gaussian fit to each residual plot.
\label{pstopstdres}}
\end{figure}

The residual distributions for the FPIX0 $p$-stop detector 
are shown in Fig.~\ref{pstopstdres}.  Each 
distribution is fitted to a Gaussian.  At small angles 
the residual distributions are non-Gaussian, because 
single pixel clusters are the dominant multiplicity and 
have a flat residual distribution. 
In addition there are 
non-gaussian ``tails'' due to the emission of $\delta$-rays, that are
discussed below. Nonetheless, the Gaussian standard deviations provide 
a commonly used measurement of the spatial resolution, that gives a
reasonable parameterization of the dominant component of the residual distribution.

\begin{figure}
\centerline{ \epsfig{figure=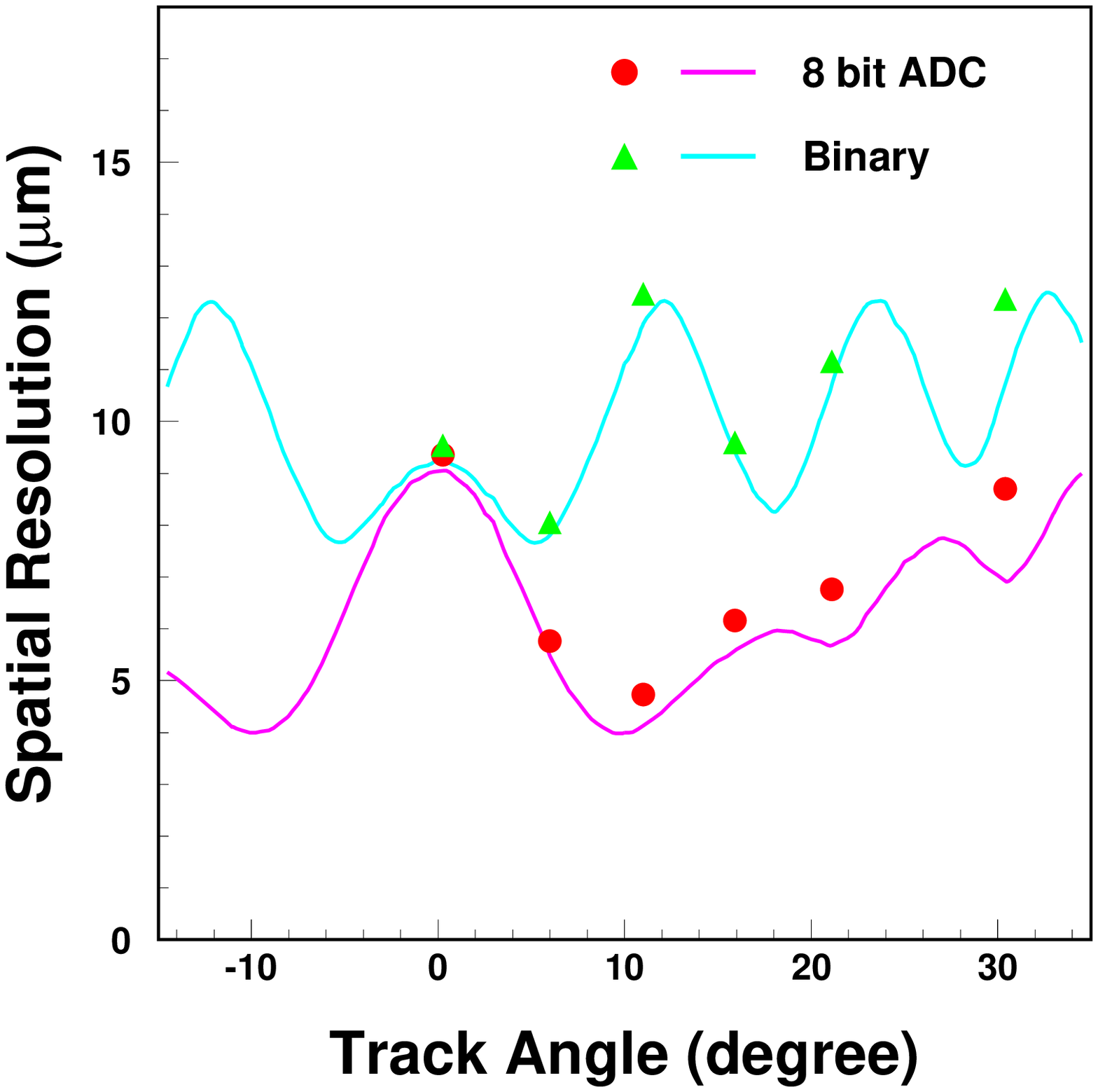, width=4in} }
\caption{ Position resolution as a function of beam incidence angle for
 the FPIX0 instrumented CiS $p$-stop sensor.
 The curves represent the predicted
 resolution Q$_{thres}$ = 2500 e$^-$. The oscillating curve is the simulated digital
 resolution; the lower curve assumes 8-bit charge digitization.  The solid circles
 are the Gaussian $\sigma$'s of the residual distribution, and the
 triangles are the $\sigma$'s extracted from fits to residual distributions made
 without using charge sharing information.
\label{res_fpix0}}
\end{figure}

Fig.~\ref{res_fpix0} shows the resolution as a function of
angle. Two curves and data points are included: the solid line and
circles show prediction and measurements 
done with an external 8 bit ADC; the dashed curve and triangular data points 
illustrate the results obtained using only $x_D$, the binary reconstructed
position, to simulate digital readout.
Note the excellent agreement between
simulation and data. The clear advantage of the analog
readout is also evident. The binary interpolation is less
accurate and features pronounced oscillations in the spatial
resolution as a function of the track incident angle.  

\subsection{Resolution versus bias voltage and readout threshold}
\begin{figure}
\begin{center}
\begin{tabular}[t]{cc}
   \epsfig{figure=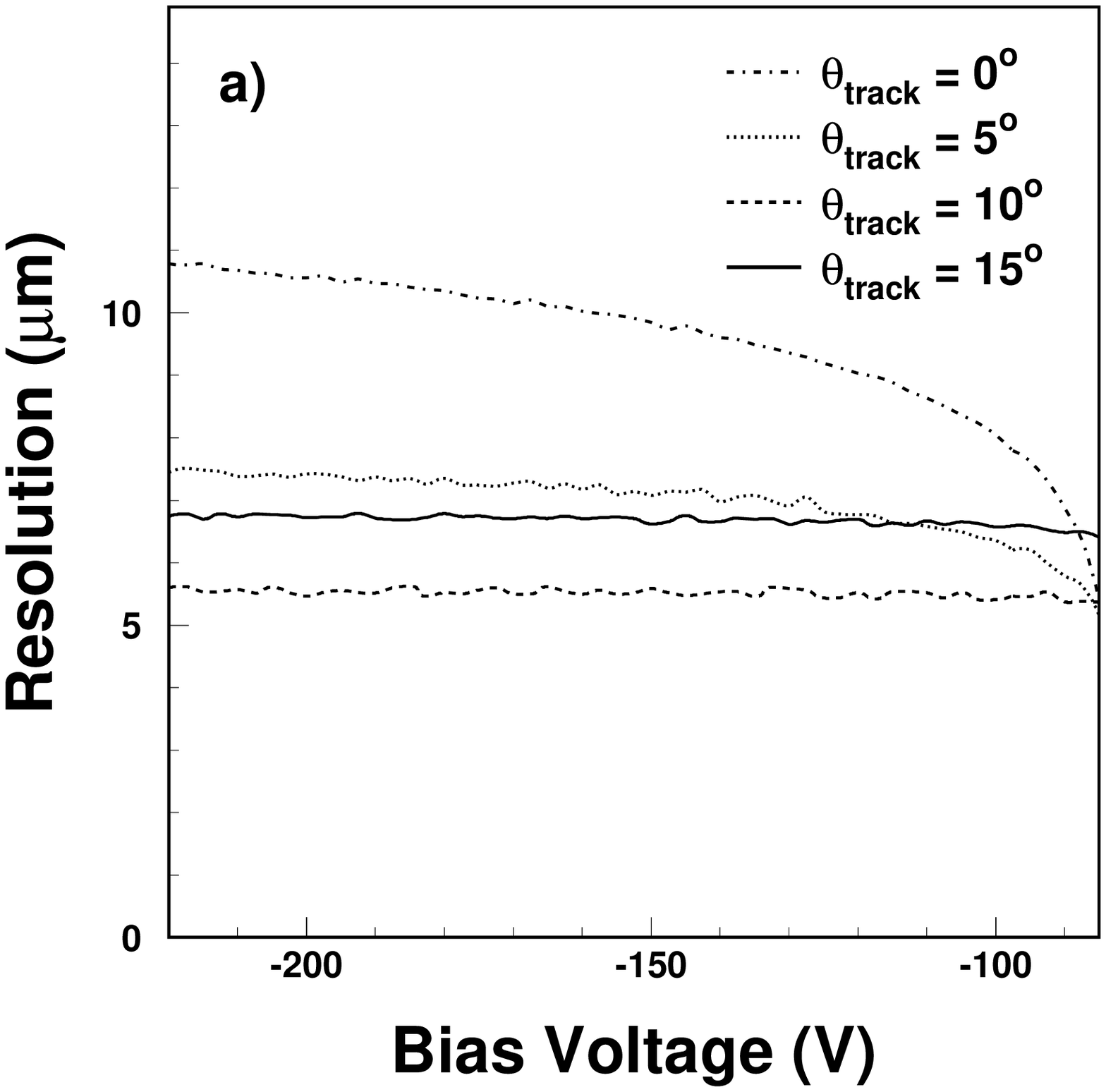,height=2.5in} &
   \epsfig{figure=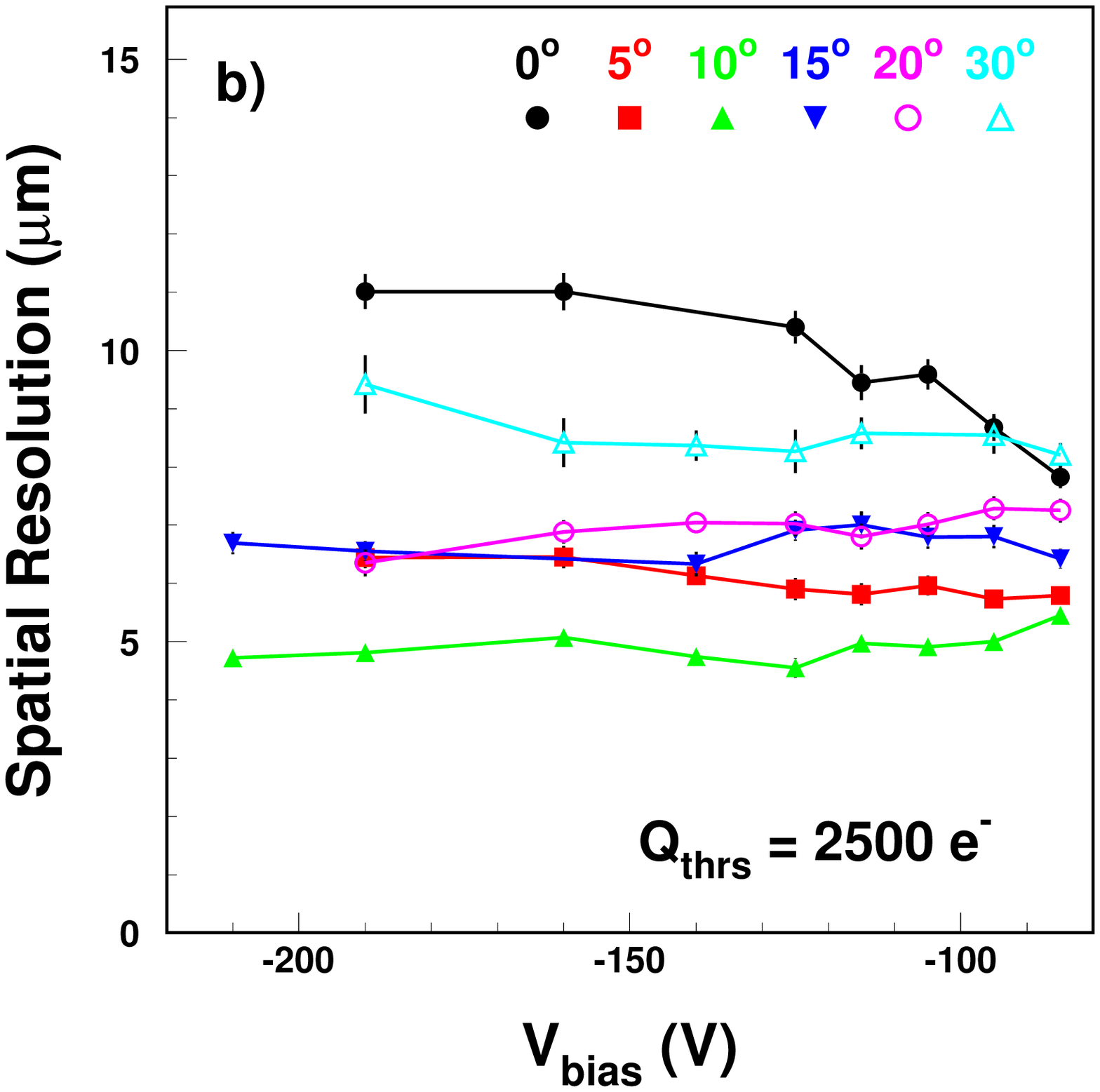,height=2.5in}
\end{tabular}
\end{center}
\caption{Spatial resolution as a function of bias voltage for (a) Monte Carlo
simulation 
and (b) measurement.
\label{bias}}
\end{figure}

Data were taken with a variety of sensor bias voltages
and readout thresholds. 
Fig.~\ref{bias}.a shows the
predicted sensitivity of the spatial resolution on bias voltage for a $p$-stop detector bump
bonded to an FPIX0 chip with 8 bit analog readout. Fig.~\ref{bias}.b shows
the corresponding data. A noticeable improvement in the resolution is obtained at small track angle
when the reverse bias is close to the depletion voltage. This is because the
longer collection time with lower bias voltage
allows more charge sharing, thus reducing the
percentage of the single pixel clusters. 

\begin{figure}
\begin{center}
\epsfig{figure=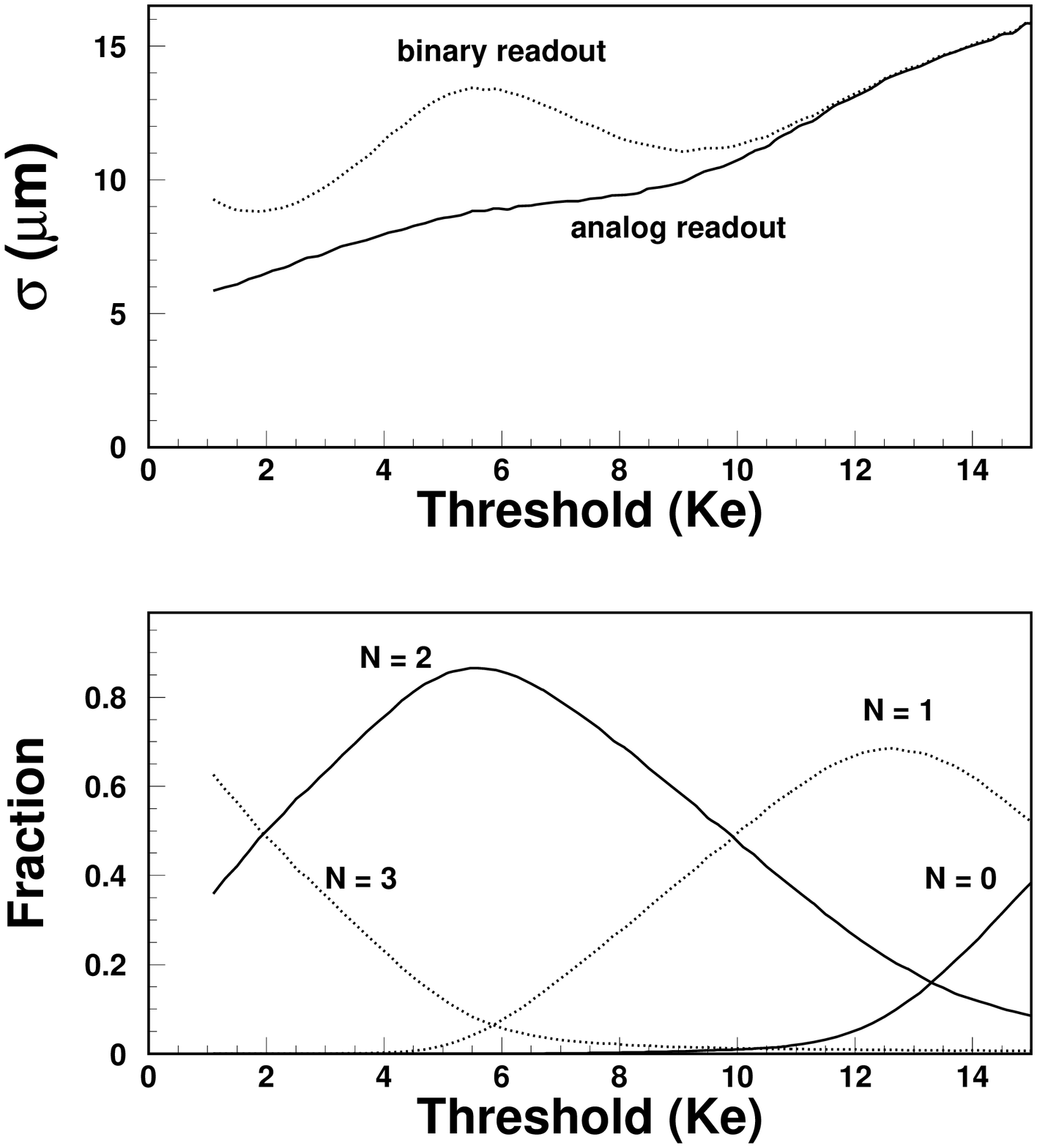,width=4in}
\end{center}
\caption{Influence of threshold on the spatial resolution (top), and 
cluster size for tracks at 0.3 rad incidence angle (bottom).
\label{resth}}
\end{figure}

\begin{figure}
\begin{center}
\epsfig{figure=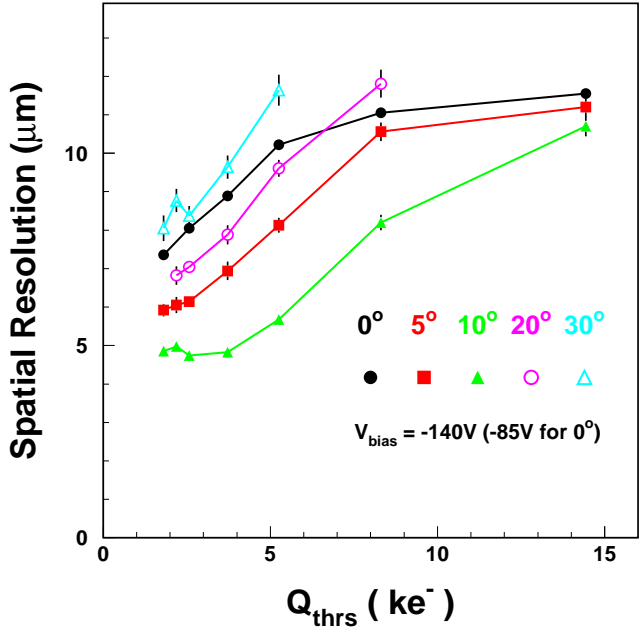,width=4in}
\end{center}
\caption{Measured spatial resolution for various 
electonic thresholds and incidence angles.
\label{resth_meas}}
\end{figure}

The discriminator threshold is one of the front-end electronics parameters
that has considerable influence on the spatial resolution. 
Fig.~\ref{resth} (top) shows  the predicted effect of increasing the 
threshold for an incident angle $\Theta$ of 300 mrad, for analog and digital 
readout.  The bottom plot shows the fraction of events having $N$ 
pixels hit for a given threshold. 
For instance, for a threshold of 2000 electrons, about 50\% of the events 
have 3 pixels hit, and about 50\% have 2 pixels hit. 
As the digital clustering algorithm
exploits the information provided by the number of pixels in a cluster, its
accuracy is best when there is an almost equal population in two different
cluster sizes: one cluster size corresponding to a track incident close to the
pixel center and the other corresponding to incidence close to the
boundary between two pixels. 
In the analog readout case, the accuracy of any position reconstruction
algorithm is degraded as the threshold increases. Note that at increasing
thresholds the efficiency becomes smaller, as illustrated by the curve
N=0, corresponding to no pixel firing because all the charge signals are
below threshold. Fig.~\ref{resth_meas} shows the corresponding measured data points
for FPIX0 hybrid detectors.

\subsection{Effective depletion depth}
The effective depletion depth of the detector can be measured from data
at large track angle. We used  two different methods.
The first method has been originally proposed by ATLAS\cite{atlas-t} 
to measure the effective depletion depth of irradiated and non-irradiated
sensors. It consists of a track-pixel position correlation estimator, 
where for each pixel over threshold whose center is at position  x$_{i}$ the 
distance z$_{i}$ between the track and the backplane is calculated with the
formula:
$z_{i}=(x_{i}-x_{inc})\times\tan{\theta}$,
where x$_{inc}$ is the x coordinate of the entrance point
extrapolated by the fitted track and $\theta$ is the track incidence angle.
The  z$_{i}$  distribution is flat between approximate 0 and the effective 
depletion depth.
The locations of rising and falling edges vary with different electronic 
thresholds.
In order to estimate the depletion depth, a detailed MC simulation is needed.

The second method consists of a length-charge correlation estimator
using the pixel charge within a cluster.
For clusters including $N\ge 3$ pixels, we can estimate an effective
depletion depth
$z_{D}=(1+\frac{q_{L}+q_{R}}{q_{middle}})\frac{N_{middle}p}{{\tan{\beta}}}$,
where q$_{L,R}$ is the charge collected by the pixel at the left (right) edge,
q$_{middle}$ is the sum of the charge collected by the $N-2$ middle pixels, 
and $p$ is the pitch along $x$.
The distribution of the estimator z$_{D}$ is peaked around the
effective depletion depth.

\begin{figure}
\begin{center}
\epsfig{figure=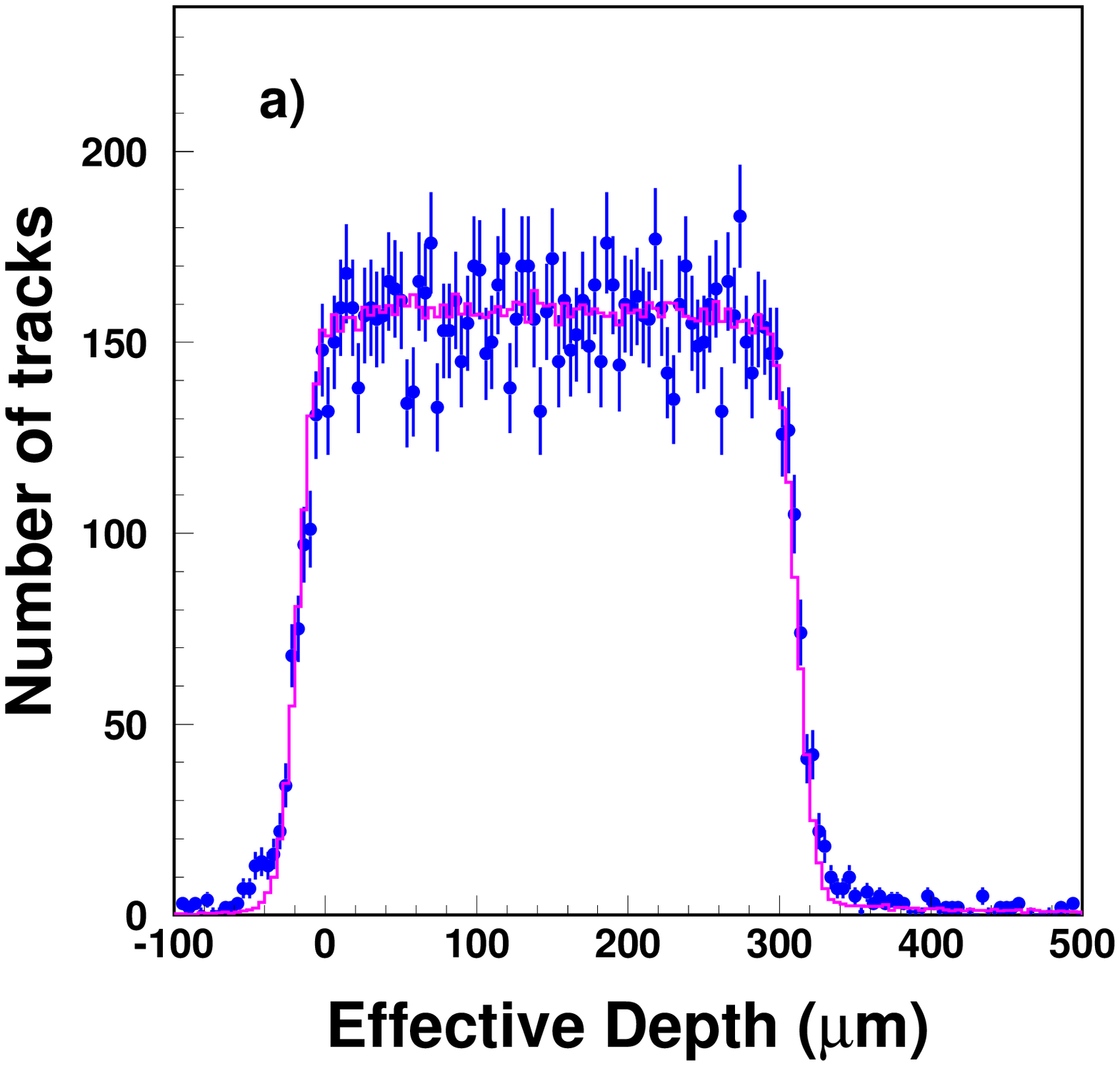,width=2.5in}
\epsfig{figure=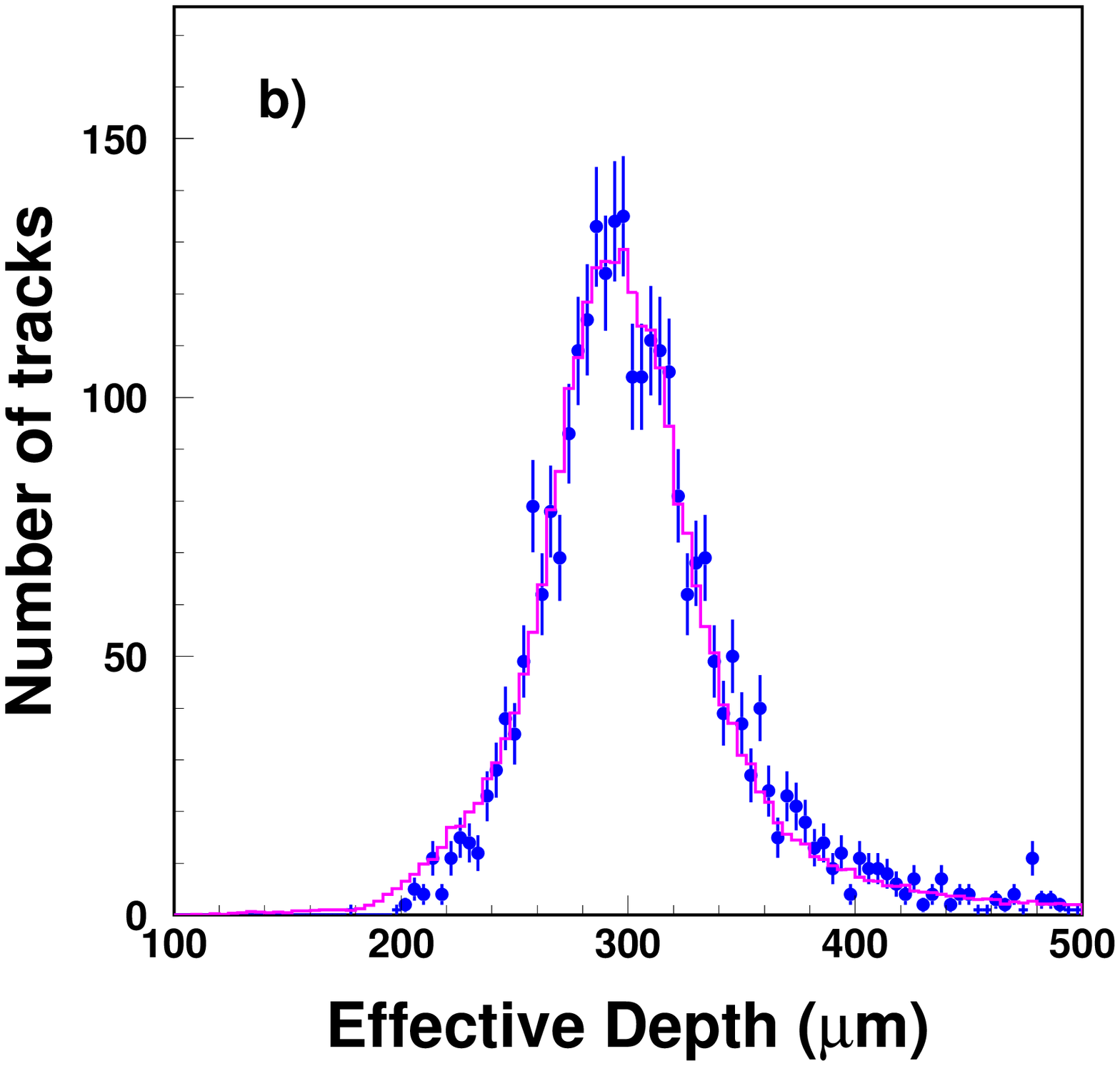,width=2.5in}
\end{center}
\caption{The effective depletion depth distribution for the CiS-FPIX0
$p$-stop detector using a) method 1, and b) method 2 as discussed in the text.
The dots are measurements and curves are MC simulation.
\label{effectivedepth}}
\end{figure}

Fig.~\ref{effectivedepth} shows the data and MC simulation for the
CiS $p$-stop-FPIX0 hybrid detector.  
Using $\beta=30^o$, the effective depth is found to be $293. \pm 1.2 \mu$m
 and $305. \pm 1.0 \mu$m for the two methods respectively. 
The difference between two
methods ($12.3 \mu$m) can be used as an estimate of the uncertainties in the
 two methods.
Note that this error corresponds to about 1$^{\circ}$ uncertainty in the track
incident angle. So the depletion depth is 
$(300 \pm 1 \pm 13)\mu$m.

\subsection{High multiplicity clusters}
\begin{figure}
\begin{center}
\epsfig{figure=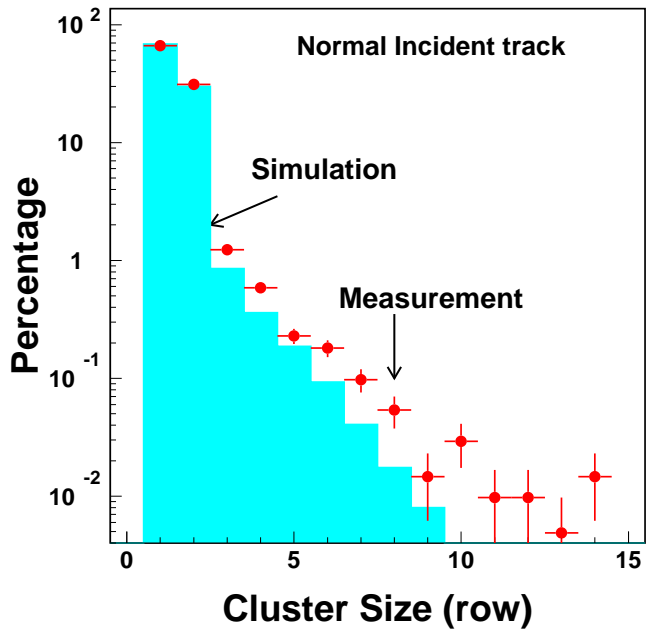,width=4in}
\end{center}
\caption{The cluster multiplicity fraction in term of number of rows for
FPIX0-$p$-stop detector at normal incidence.
The points are data and shaded histogram is the Monte Carlo prediction.
\label{multiplicity} }
\end{figure}

A small fraction of events 
have clusters with a number of hits 
greater than expected, given only the track trajectory and diffusion.
Fig.~\ref{multiplicity} shows the measured frequency  of high-multiplicity 
clusters for tracks at normal incidence. Our Monte
Carlo simulation, which includes energetic $\delta$-rays, reproduces
the measured cluster multiplicity reasonably well.

\subsection{Resolution function shape}
The pixel residual distribution (or resolution function) 
deviates from a Gaussian in two ways:  the first effect is produced by
the non-optimal
the charge sharing at almost normal incidence that produces 
a large fraction of 1 pixel
clusters; the second is related to large cluster size events due to delta-ray
emission.
These effects have been alluded to before and are now discussed
quantitatively
using a sample of tracks at nominal normal incidence.

\begin{figure}
\begin{center}
\epsfig{figure=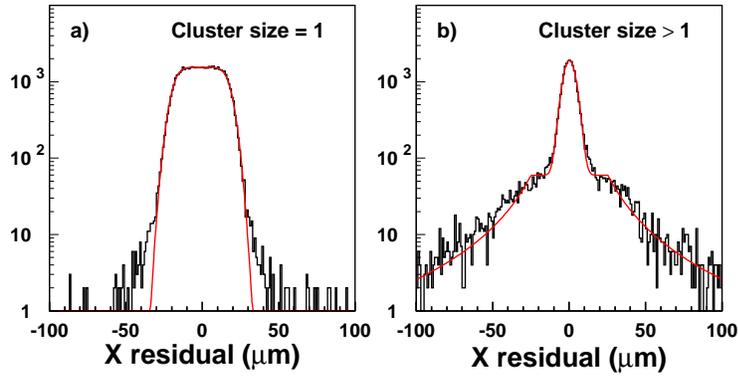,width=4in}
\end{center}
\caption{ Residual distributions for FPIX0 $p$-spray detector at zero degree.
 The plot on the left shows the distribution for cluster size 1 and the plot
 on the right for clusters with multiplicity higher than 1.
\label{boxandtail}}
\end{figure}

Fig.~\ref{boxandtail} shows the residual distributions for the FPIX0
$p$-spray detector taken with the beam nominally at normal incidence.  The 
plot on the left shows the residual for one-pixel clusters.  
The plot on the right shows the residual for clusters of two or
more pixels, with a fit with the function $F(x)$:
\begin{equation}
 F(x)=N_1 F_{G}(x)+ N_2 F_{NG}(x),
\label{resfun}
\end{equation} 
\noindent
where $F_{G}$ is a Gaussian, and $F_{NG}$ is defined as:
\begin{equation}
 F_{NG}(x)= \left\{
 \begin{array}{c}
 \frac{A_{pl}}{|r_{c}|^{\gamma}}\ {\rm for}\ |x|<r_{c}\\
 \frac{A_{pl}}{|x|^{\gamma}}\ {\rm for}\ |x|>r_{c}
 \end{array}
 \right.,
\label{powerlaw1}
\end{equation} 
\noindent
$A_{pl}$ is a normalization constant, $r_{c}$ is the half width 
of the constant term, and $\gamma$ is the exponent of the power-law dependence.
The non-Gaussian fraction
$N_2/(N_1+N_2)$ accounts for 18\% of the total number of entries in the
distribution.

 The naive expectation of a
rectangular shape 50 $\mu$m for the residual distribution 
for one pixel clusters is distorted by two effects. The population
at the ``edges'' of the rectangle is depleted by diffusion, inducing two
pixel clusters for track incident near the pixel periphery. On the other 
hand, a broadening of this ideal distribution is induced by
threshold dispersion and electronic noise, as well
as the small fraction of mismeasured extrapolated position $x_T$.

\begin{figure}
\begin{center}
\epsfig{figure=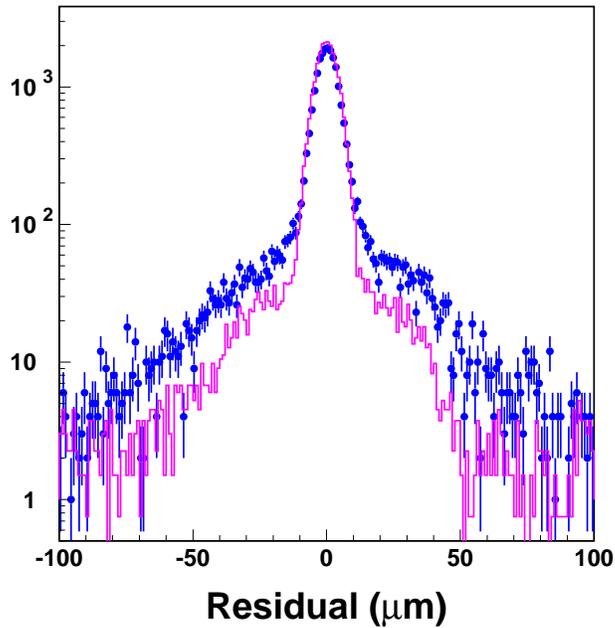,width=4in}
\end{center}
\caption{ Residual distributions for FPIX0 $p$-spray detector at zero degree.
Points represent data for tracks at normal incidence for several different
 runs and the curve represents our simulation predictions. Both distributions
include  only clusters with multiplicity greater than 1.
\label{deltaray}}
\end{figure}

The second factor that makes the pixel residual distributions non-Gaussian 
is $\delta$-ray emission. 
Low energy $\delta$-rays which stop in one of the pixels crossed by the
particle skew the charge sharing and degrade the resolution.  Higher energy
$\delta$-rays cross one or more pixel boundaries and distort the position
measurement even more. We have studied this effect with the Monte Carlo
simulation described before. Fig.~\ref{deltaray} shows a comparison between
predictions and data. The
distributions are shown using a log scale to show the tails of the residual
distribution more clearly. The simulation accounts for about 1/2 of the broad
component of the residual distribution. This may be in part due to
instrumental effects such as the charge losses near the pixel boundaries of
the $p$-spray devices.

\subsection{Charge sharing across columns}
Charge-sharing between columns occurs for less 
than $5\%$ of the tracks in our data sample.
We collected enough data
to study this effect only at 0$^{\circ}$, where
the diffusion is the basic sharing mechanism. 

\begin{figure}
\begin{center}
\epsfig{figure=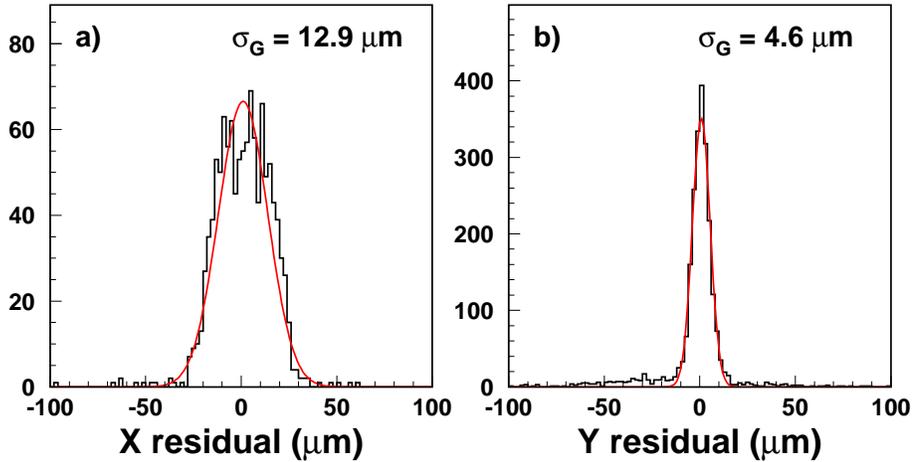,width=5in}
\end{center}
\caption{
The plots show residual distributions in x (short pixel dimension)
and y (long pixel dimension) for events where the charge-sharing
occurs across 2 columns but charge is collected in a single row.
\label{sharingcol} }
\end{figure}

For $p$-stop sensors the effective charge sharing region is a rectangular ring 
with approximately uniform thickness along the pixel boundary. 
Fig.~\ref{sharingcol} shows the residual distributions $\sigma _x$
and $\sigma_y$  of 1 row and 2 column clusters.
The measured $y_p$ is calculated with the procedure described before,
using a linear eta correction determined from the data.
The spatial resolution along $y$ is consistent with the $x$ resolution when charge
sharing allows for interpolation between the pixel centers.

\section{Conclusions}
An extensive test beam study of several hybrid pixel detectors 
has demonstrated that both the sensors and the front-end electronics
chosen for BTeV perform according
to expectations. The charge collection properties of the sensors studied are 
well understood. We have shown that 2-bit analog information is 
satisfactory and,
consequently, that the final version of the front-end electronics, featuring a
3-bit flash ADC will provide the excellent spatial resolution needed
to achieve the BTeV physics goals.

\section{Acknowledgements}
We would like to thank Fermilab for providing us with the dedicated beam
time for our test and the excellent infrastructure support. We are grateful
to the ATLAS collaboration, with special thanks to their pixel group for the
sensors that were used in this test and many useful discussions. We are
indebted to Colin Gay for the data acquisition software.
We also thank the US National Science Foundation and the Department of Energy
for support. The Universities Research Association operates Fermilab for the
Department of Energy.


\end{document}